\documentclass{iau} 
\usepackage{graphicx}

\title[JD 11.~~Revisiting the spectra of old AGN friends] 
{Revisiting old (AGN) friends – what’s changed in their spectral looks}

\author[Hartmut Winkler]   
{Hartmut Winkler$^1$
}

\affiliation{$^1$Department of Physics, University of Johannesburg, \\ PO Box 524,
2006 Auckland Park, Johannesburg, South Africa \\ email: {\tt hwinkler@uj.ac.za} }

\pubyear{2019}
\volume{356}  
\jname{Nuclear Activity in Galaxies across Cosmic Time}
\editors{M. Povic, J. Masegosa, H. Netzer, P. Marziani, P. Shastri, S.B. Tessema \& S.H. Negu, eds.}
\begin{document}

\maketitle

\begin{abstract}
Active Galactic Nuclei (AGN) have long been known to be variable, but the amplitude,
timescale and nature of these changes can often differ dramatically from object
to object. The richest source of information about the properties of AGN and the
physical processes driving these remains the optical spectrum. While this spectrum
has remained remarkably steady over decades for some AGN, other objects, referred
to as Changing Look AGN, have experienced a comprehensive spectral transformation.
Developments in the detection technology have enabled detailed probing in
other wavebands, highlighting for example often quite different variability patterns
for high energy emission. This paper explores the current characteristics of some
long-known (and almost forgotten) Seyfert galaxies. It compares their present optical
spectral properties, determined from recent observations at the South African
Astronomical Observatory, with those from much earlier epochs. It furthermore
considers the implication of the changes that have taken place, alternatively the
endurance of specific spectral features, on our understanding of the mechanisms of
the observed targets in particular, and on AGN models in general.
\end{abstract}

\firstsection 

\section{Introduction}

Seyfert galaxies were identified as a class in the 1940's (\cite{Seyfert43}), and have elicited astrophysical interest ever since. While initially restricted to a few, mostly nearby objects, the list of galaxies belonging to the class grew substantially in the 1970's, due to systematic surveys such as the one of Markarian (\cite{Markarian81}), and as a result of the first x-ray source identification programmes. The AGN explored in this paper, Fairall 9, IC 4329A and Mkn 926, were discovered to be Seyferts during such investigations, and their names have been amongst the most recognisable in the discipline ever since. All three have at one stage been suggested as candidates for the title of `nearest quasar'.

These AGN have now been known to exhibit broad-line Seyfert galaxy spectra for over 40 years. Over that period they have been frequently (though not regularly) re-observed, and in addition considerable data sets have also been collated in other wave bands, such as in x-rays and in the infrared. They have become suitable probes to investigate the medium-term ($\sim$20-50 year period) behaviour of luminous broad-line AGN.

Inspecting the nature of variations over the medium-term also enables one to determine the commonality of the relatively recently identified `changing look' phenomenon. This refers to a relatively rapid, dramatic change in the optical spectrum. Typically, initially strong broad lines of a changing look AGN disappear almost entirely over a few months, meaning that what was once a Seyfert 1 or 1.5 is transformed into a Seyfert 1.8 or 1.9 over a time period considered too short for major nuclear obscuration changes. Changing look AGN therefore challenge the unified AGN model that considers Seyfert classes to merely be products of different inclination angles of a dusty torus constituting the outer parts of the nucleus (see, e.g., the discussion in \cite{Oknyansky19}).

\section{Observations}

This paper discusses optical spectroscopic observations carried out on the night of 9 Jul 2019 with the SpUpNIC spectrograph (\cite{Crause19}) mounted on the 1.9~m telescope at the South African Astronomical Observatory in Sutherland. Two 1200~s integrations were made for each object, bracketed by 10~s Ar wavelength comparison spectra immediately before and after each AGN spectrum. The grating employed was one giving a low resolution and resulting in a useful spectral wavelength range 3250-8500 \AA, and the slit width was set to 2.7 arcsec on the sky. The flux calibration was achieved by means of a spectrum of the standard star LTT 3864. The processing of the frames included standard bias subtractions, flatfield corrections and the cleaning of pixels affected by cosmic rays.


\section{Fairall 9}

Fairall 9 was identified as a quasar-like emission-line object during a spectral survey of galaxies with compact nuclei (\cite{Fairall77}). \cite{Ward78} independently identified the object as the optical counterpart of the x-ray source 2A 0120--591, and published the first detailed spectrum displaying strong broad lines. Confirming the spectrum and presenting photometric measurements indicating $V \sim 13.2$, $B-V \sim 0.2$ and $U-B \sim -1.0$, \cite{West78} classified the object as a quasar.

From about 1981, Fairall 9 experienced a remarkable drop in luminosity. By 1984, the Balmer lines had weakened to about 20\% of their 1981 values relative to the forbidden lines (\cite{Kollatschny85}, \cite{Wamsteker85}). The luminosity decline, and the subsequent slow partial rebrightening, was also closely monitored in the ultraviolet and infrared (\cite{Clavel89}). Only minimal spectral changes were detected in 1987-1988 (\cite{Winkler92}). A 1993 spectrum displays a rather weak broad component, with a red shoulder now quite clearly distinguishable (\cite{Marziani03}). In 1994, a large reverberation mapping campaign determined a broad-line radius of 23 days (\cite{SantosLleo97}).

Thereafter, spectral data was secured a lot less frequently as researchers moved to study other AGN. While a lot of data has still been collected for Fairall 9 post-1994, very little of it covers the optical regime. The optical spectral behaviour over this lengthy period can therefore only be inferred from what was measured in other wave bands. There is no evidence of brightening comparable to the levels witnessed in the late 1970's, neither does anything suggest that the AGN `switched itself off' at any stage.

\begin{figure}[h]
\begin{center}
 \includegraphics[width=5.4in]{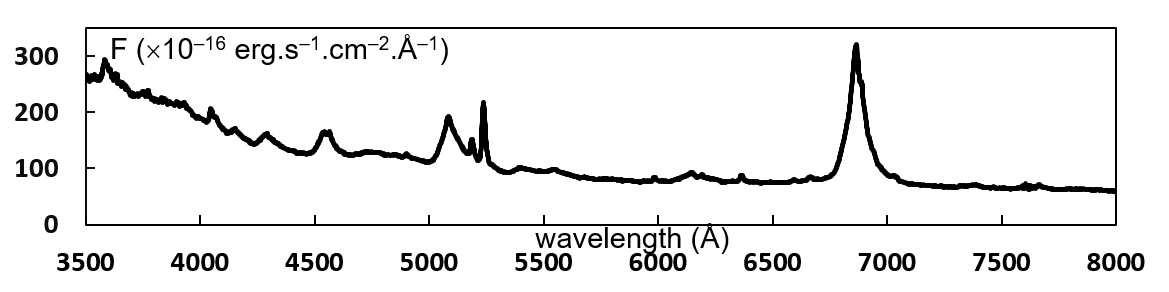} 
 \caption{Optical spectrum of Fairall 9 on 9 July 2019.}
   \label{fig1}
\end{center}
\end{figure}

The July 2019 spectrum of Fairall 9 is illustrated in Fig.\,\ref{fig1}. It looks remarkably similar to the spectrum from the late 1980's shown in \cite{Winkler92}, when the luminosity of Fairall 9 had strengthened to intermediate from the 1985 minimum. There is no more obvious sign of the `red shoulder', and the FeII emission bands also resemble the same lines in the corresponding earlier spectra.

\section{IC 4329A}

The AGN nature of IC 4329A, an edge-on galaxy with a striking dust lane (see, e.g. \cite{Malkan98}), was established in 1972 (\cite{Disney73}). The discovery paper highlights the strong broad-line spectrum, with Balmer line zero intensity widths of up to 13000 km.s$^{-1}$. It furthermore records an H$\alpha$-to-H$\beta$ ratio of over 8, a clear sign of significant nuclear reddening and extinction. The high Balmer-line flux ratios were confirmed in many later studies (e.g., \cite{Wilson79}), although it is now also clear that the intrinsic ratio may be somewhat greater than was initially assumed, meaning that the intrinsic nuclear luminosity is not high enough to warrant classification as a quasar.

The discovery spectrum also revealed a secondary peak (red shoulder) of the broad H$\beta$, corresponding to a rest wavelength of 4900 \AA, and a recession velocity of roughly 2300 km.s$^{-1}$ relative to the primary H$\beta$ peak. Significantly, this secondary peak also appears in varying degrees in many later spectra (e.g., \cite{Winkler92}, \cite{Winge96}).

\begin{figure}[h]
\begin{center}
 \includegraphics[width=5.4in]{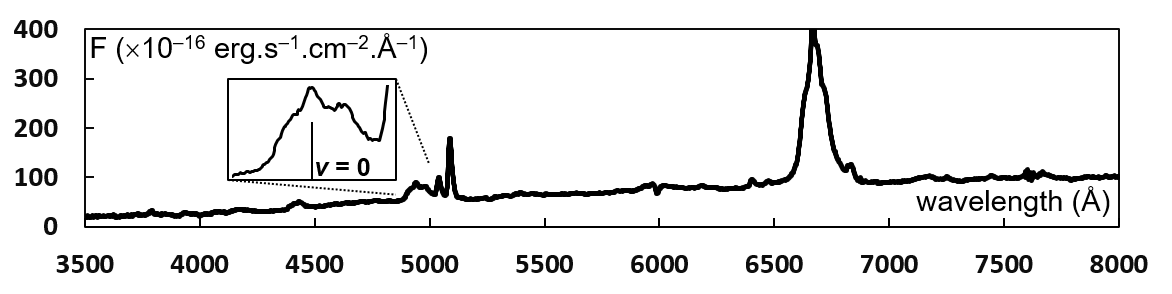} 
 \caption{Optical spectrum of IC 4329A on 9 July 2019, including an enlarged display of the H$\beta$ profile.}
   \label{fig2}
\end{center}
\end{figure}

Figure \,\ref{fig2} displays the spectrum most recently obtained for IC 4329A. It does not differ markedly from the other spectra published over the past almost 50 years. Notably, the redshifted second peak of the H$\beta$ emission line is still clearly distinguishable, implying that its presence is associated with a stable nuclear configuration that the AGN returns to frequently. It is pointed out that there are spectra where the red peak appears absent (e.g. \cite{Morris88}).

\section{Mkn 926}

Mkn 926, also frequently referred to as MCG --2-58-22, was identified as the optical counterpart of the bright x-ray source 2A 2302--088 by \cite{Ward78}. \cite{Morris88} show a spectrum recorded in 1984, when the object was in a luminous state. The broad-line component weakened substantially from about 1987 (\cite{Winkler92}), and a further significant broad-line luminosity decline ensued in around 1993 (\cite{Kollatschny06}). Mkn 926 was also observed spectroscopically during the Sloan Digital Sky Survey (\cite{Ahn14}), and throughout the early years of the present century the Balmer broad lines remained weak (consistent with a Seyfert 1.8 classification), though with a complex profile that included a distinct red saddle at $\sim10000$ km.s$^{-1}$.

\begin{figure}[h]
\begin{center}
 \includegraphics[width=3.4in]{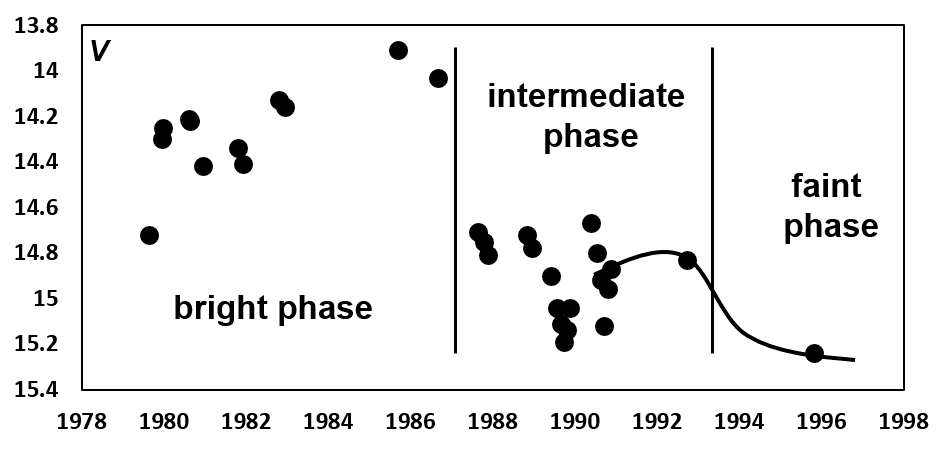} 
 \caption{$V$ light curve of Mkn 926 from 1978-1996. Magnitudes from \cite{Doroshenko81}, \cite{Mallama83}, \cite{deRuiter86} (converted to $V$-magnitudes using the conversion equations given in that paper), \cite{Hamuy87}, \cite{Winkler92b}, \cite{Winkler97}, and converted to a 10 arcsec aperture diameter. The curved line estimates the magnitudes for 1990-1995 from spectral continuum decline determined by \cite{Kollatschny06}.}
   \label{fig3}
\end{center}
\end{figure}

The evolution of Mkn 926 since its discovery is illustrated in Fig.\,\ref{fig3} in the form of an approximate $V$-magnitude light curve. Three distinct phases can be distinguished: i) a bright phase lasting to late 1986, when the object was one of the more luminous AGN in the not-too-distant ($z < 0.05$) universe, ii) and intermediate phase lasting to about 1993 where Mkn 926 was much less luminous than before, but still exhibited typical broad-line Seyfert characteristics and medium-term luminosity fluctuations, and iii) a faint phase from about 1994, where the spectrum is dominated by narrow lines (including comparatively strong lines associated with low-ionisation), though a weak broad Balmer line spectrum always remains present. 

\begin{figure}[h]
\begin{center}
 \includegraphics[width=5.4in]{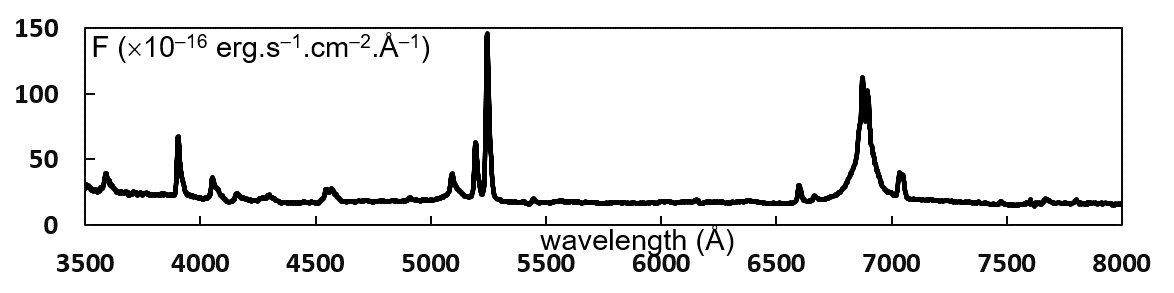} 
 \caption{Optical spectrum of Mkn 926 on 9 July 2019.}
   \label{fig4}
\end{center}
\end{figure}

The latest spectrum of Mkn 926 is shown in Fig.\,\ref{fig4}. The broad component is now even weaker than it was around the time of the SDSS observations, and the `red saddle' evident in some earlier spectra is no longer visible. The broad component of H$\beta$ is only detectable because of the high S/N, and a spectrum of lower quality might have led to Mkn 926 being described as a Seyfert 1.9.

While the Balmer and helium emission in the current spectrum is dramatically lower than 40 years ago, indeed so different that Mkn 926 can be classified as a changing look AGN, the object's peculiar narrow line spectrum has not undergone any obvious change. In particular, the comparatively strong [O II] and [S II] lines are more reminiscent of a LINER, and this may in some way relate to this AGN's peculiar spectral evolution.

\section{Discussion and Conclusion}

Three well-known, comparatively nearby Seyfert 1 galaxies, Fairall 9, IC 4329A and Mkn 926, all of which were studied extensively spectroscopically in the last quarter of the 20th century, have been reobserved. The spectra of Fairall 9 and IC 4329A were found to closely match spectra from 30 years ago. Fairall 9 has been significantly brighter (and fainter) in some earlier epochs, which suggests that comparatively stable states do exist that the AGN returns to after its bright phases. While the nuclear luminosity of IC 4329A appears comparatively stable, some line profile changes have been evident. In particular, an emission peak 2300 km.s$^{-1}$ redward of the H$\beta$ peak, suggestive of Keplerian motion in the nuclear regions and seen in spectra from $\sim 25$ years ago, remains visible.

In contrast to the other two AGN investigated here, Mkn 926 has shown no signs of re-entering the brighter states it was in from the mid-1977s to the early 1990s. If there are cycles of activity, the timeframe of this cycle is at least 50 years long. Its semi-quiescent state has now endured for about 25 years. In all cases, the medium-term spectral evolution is determined by physical conditions inside the nucleus, and as such constrains models to explain the physical mechanism driving the AGN.




\end{document}